\documentclass[twocolumn]{IEEEtran}
\usepackage{url,amsmath,graphicx}
\usepackage{amsmath,amssymb,amscd,latexsym,dsfont,amsthm}
 \usepackage{stfloats}
 \usepackage{eqparbox}
  \usepackage{array}
  \usepackage{fixltx2e}
  \usepackage{stfloats}
   \usepackage{cite}
   \usepackage{url}
\usepackage{cite}
\usepackage{graphicx}
\usepackage{psfrag}
\usepackage{url}
\usepackage{stfloats}
\usepackage{array}
\usepackage{fancyhdr}
\usepackage{psfrag}
\usepackage{subcaption}
\usepackage{graphicx}
\usepackage{lipsum}
\usepackage{amssymb}
\usepackage{color}
\begin{document}
%
\title{ Terahertz  User-Centric Clustering  in the Presence of  Beam Misalignment}

\author{ Khaled Humadi,  \textit{ Member, IEEE}, Im\`ene~Trigui,  \textit{Member, IEEE}, \\Wei-Ping Zhu, \textit{Senior Member, IEEE}, and  Wessam Ajib, \textit{Senior Member, IEEE}. 

%
\thanks{

This work was supported by the Regroupement
Strategique en Microsystemes du Quebec (ReSMiQ) and the Fonds de recherche
du Québec. 

Khaled Humadi is with the Department of Electrical Engineering, Polytechnique Montréal, Montréal, QC
H3T 1J4, Canada, e-mail: khaled.humadi@polymtl.ca, Wei-Ping Zhu is with the Department of Electrical and Computer Engineering, Concordia University, Montreal, QC H3G 1M8, Canada, e-mail: weiping@ece.concordia.ca. Im\`ene~Trigui and Wessam Ajib are with the Departement d’informatique, Université du Québec à Montréal, Montreal, QC H2L 2C4, Canada, e-mail: trigui.imen@courrier.uqam.ca and ajib.wessam@uqam.ca.}
\vspace{-0.3cm}
}
\maketitle

\begin{abstract}
Beam misalignment  is one of the main 
challenges for the design of reliable  wireless systems
in terahertz (THz) bands.
This paper investigates how to apply user-centric base station (BS) clustering as a valuable  add-on in THz networks. In particular, to reduce the impact of beam misalignment,  a user-centric BS clustering design that provides multi-connectivity via  BS cooperation is investigated.   The coverage probability is derived  by leveraging an accurate approximation of the aggregate interference distribution that captures the effect of beam misalignment and THz fading.  The numerical results reveal the impact of beam 
misalignment with respect
to crucial link parameters, such as
the transmitter’s beam width and  the serving  cluster size, demonstrating that user-centric BS clustering is a promising enabler of  THz networks.

\end{abstract}

\begin{IEEEkeywords}
User-centric clustering, Terahertz, system coverage, beam misalignment, BS cooperation, stochastic geometry.
\end{IEEEkeywords}

%
\IEEEpeerreviewmaketitle
\vspace{-0.5cm}
\section{Introduction}

Terahertz (THz) band promises to provide very broad bandwidth that is  required to support the unprecedented  high data rate applications \cite{elayan2018terahertz}, \cite{elayan2019terahertz}.  In order to fully realize the benefits of THz band, transceivers must be supported with  directive antennas   to help compensate
for substantial propagation losses  due to free space attenuation and molecular absorption.  However, the high antenna directivity  requirements in the THz connection  
makes it more vulnerable to beam alignment errors, which significantly degrades its performance. \cite{boulogeorgos2021directional},\cite{ghafoor2021next}. 
As a result, developing efficient beam alignment techniques is one of the main challenges  in THz wireless networks. 

Recently, a considerable amount of  research was dedicated to  analyze and mitigate the effect of beam misalignment in wireless networks operating at higher frequency bands such as millimeter wave (mmWave) and  THz bands   \cite{papasotiriou2020performance},\cite{badarneh2022performance}. 
Moreover, non-orthogonal multiple access (NOMA)  has been applied in \cite{ding2022design} to overcome the effect of 
antenna beam misalignment in THz wireless systems. The analytical results developed in \cite{ding2022design} illustrate that 
using NOMA can efficiently enhance the system coverage performance and user connectivity. In \cite{papasotiriou2022analytical} the technique of deploying  reconfigurable intelligent surfaces (RIS) is used  as a promising enabler to
reinstate the interrupted THz links. Specifically, the combined use of directive beams for mitigating the pathloss and RISs for
restoring the line-of-sight is analyzed.
Unlike these existing works focusing on the use of RIS and NOMA in THz, this paper investigates how to use BS clustering and cooperation  as
a type of add-on in THz wireless networks. In particular,  to reduce the impact of beam misalignment,
macro diversity via BS cooperation is deployed where joint
transmission from a cluster of serving BSs enhances  communication reliability. 

In the absence of beam misalignment, only few works investigated the performance of user-centric THz wireless networks  \cite{humadi2021coverage},\cite{humadi2022user}. In particular,  \cite{humadi2021coverage} compared  both adaptive and static  BS clustering designs in terms of coverage probability improvement in THz networks and provided the necessary analytical performance models for both clustering approaches. A more comprehensive framework  was also presented in \cite{humadi2022user} where user-centric clustering is applied in  multi-tier networks where  mmWave and THz clusters are opportunistically selected to serve a typical user.  However, none 
of the existing studies has addressed the unique issue associated with THz communications, namely improper alignment.

This paper fills the gap   by integrating 
user-centric BS clustering into THz networks  that suffer from significant array gain variation and even
signal outage due to beam misalignment. In order to ensure high-reliability, we implement an adaptive BS clustering for the THz user-centric wireless network and study the effect of beam misalignment on the system coverage performance.
The obtained results demonstrate that user-centric BS cooperation can significantly enhance the overall system coverage
performance  of THz networks in the presence of beam misalignment errors.

 \vspace{-0.35cm}
\section{System Model}\label{System Model}
\subsection{Network Model
}
Consider a downlink network where THz BSs are densely deployed to serve randomly distributed users. The spatial distributions of  Users and BSs are modeled as 
 two independent homogeneous Poisson point processes (PPPs) defined as $\Phi_u=\{x_i\}\in\mathbb{R}^{2}$ and $\Phi_b=\{y_k\}\in\mathbb{R}^{2}$ with different densities $\lambda_u$ and $\lambda_b$, respectively, where $x_i$ and $y_k$ represent the spatial locations  of the  $i$-th user and  the $k$-th BS, respectively.

All BSs and users in the network are assumed to be equipped with highly directional antenna arrays. 
In order to make the  analysis tractable, we used a sectored antenna model to define the gain radiation patterns of such antenna arrays \cite{di2015stochastic}. This model describes the array gain radiation pattern using three  fundamental parameters: main-lobe gain, $G_v^M$,  back and side-lobes gain,  $G_v^S$, and the antenna beamwidth, $\theta_v$, where $v\in \{b,\;u\}$ refers to the BS and user. To obtain analytically tractable closed forms for the gain pattern and antenna beamwidth, we consider uniform linear arrays with half-wavelength space between antenna elements \cite{mahafza2016radar}.
Consequentely, the effective antenna gain $G_v(\theta)$ in a given direction can be defined as\vspace{-0.2cm}
\allowdisplaybreaks
\begin{eqnarray}
     G_v(\theta)=\left\{
     \begin{array}{ll}
     G_v^M=\frac{2\pi N_v^2 \sin\Big(\frac{3\pi}{2N_{v}}\Big)}{\theta_v N^2_{v}\sin\Big(\frac{3\pi}{2N_{v}}\Big)+(2\pi-\theta_v)},  |\theta|\leq\frac{\theta_v}{2} \\\\
     G_v^S=\frac{2\pi}{\theta_v N^2_{v}\sin\Big(\frac{3\pi}{2N_{v}}\Big)+(2\pi-\theta_v)},   \textrm{otherwise} 
\end{array}
\right.
\label{GModel}
\end{eqnarray}
where $\theta$ is the angle of the boresight direction uniformly distributed in  $[-\pi,\;\pi]$,  
 $N_{v}$ denotes the number of elements in the BS/user antenna array. The  beamwidth of the antenna array, $\theta_v$,  can be obtained as\vspace{-0.3cm}
\begin{eqnarray}
\theta_v=2\arcsin\Big(\frac{2.782}{\pi N_{v}}\Big)
\label{thetaw}
\end{eqnarray}
Note that  in THz, to transmit/receive multiple beams, BSs and users are 
supported with large antenna arrays consisting of smaller
 subarrays. This enables each node to steer multiple beams in different directions. As such, $N_v$ refers to the number of antenna elements in each subarray.

\vspace{-0.35cm}
\subsection{Beam Alignment Models}
To take into account the effect of  beam misalignment, we adopt the truncated
Gaussian model  \cite{di2015stochastic}. This model counts for both the antenna beamwidth and the standard deviation of the arrays beam alignment 
error. As such, the probability of  alignment error in the desired direction is given by\vspace{-0.2cm}
\begin{eqnarray}
p_{e,v}(\theta_v,\sigma_v)&=&Pr\Big[\big|\Xi_{v}\big|> \frac{\theta_v}{2}\Big]\nonumber\\
&&\!\!\!\!\!\!\!\!\!\!\!\!\!\!\!\!\!\!\!\!\!\!\!\!\!\!\!\!\!\!\!\!\!\!\!\!\!\!\!\!\!\!\!\!=1-\bigg(1-2Q\Big(\frac{\theta_v}{\sqrt{2\sigma_v^2}}\Big)\bigg) \bigg(1-2Q\Big(\frac{\pi}{\sqrt{\sigma_v^2}}\Big)\bigg)^{-1}
\label{beamError}
\end{eqnarray}
where $v\in\{b,\;u\}$, $\Xi_{v}\sim{\cal N}(0,\sigma_v^2):$ $\Xi_{v}\in [-\pi,\;\pi]$ is the beam alignment error, and $\sigma_{v}$ denotes the error standard deviation. 
Using the gain model in (\ref{GModel}) and the beam alignment error in (\ref{beamError}), the total gain of a link between the typical user and a serving BS can be modeled as a discreet  random variable (r.v.)  with  probability mass function (pmf) as follows.
\begin{eqnarray}
G(\theta_b,\theta_u,\sigma_b,\sigma_u)&=&\nonumber\\
&&\!\!\!\!\!\!\!\!\!\!\!\!\!\!\!\!\!\!\!\!\!\!\!\!\!\!\!\!\!\!\!\!\!\!\!\!\!\!\!\!\!\!\!\!\!\!\!\!\!\!\!\!\!\!\!\!\left\{\begin{array}{ll}
     G_{1}=G_{b}^MG_{u}^M, \;\;\textrm{w.p}\;\;\;  p_{1}= \widetilde{p}_{e,b}(\theta_b,\sigma_b)\widetilde{p}_{e,u}(\theta_u,\sigma_u)  \\
     G_{2}=G_{b}^SG_{u}^M, \;\;\;\textrm{w.p}\;\;\;  p_{2}= {p}_{e,b}(\theta_b,\sigma_b)\widetilde{p}_{e,u}(\theta_u,\sigma_u) \\
     G_{3}=G_{b}^MG_{u}^S, \;\;\;\textrm{w.p}\;\;\;  p_{3}= \widetilde{p}_{e,b}(\theta_b,\sigma_b){p}_{e,u}(\theta_u,\sigma_u)\\
     G_{4}=G_{b}^SG_{u}^S, \;\;\;\;\textrm{w.p}\;\;\;  p_{4}= {p}_{e,b}(\theta_b,\sigma_b){p}_{e,u}(\theta_u,\sigma_u),
\end{array}
\right.
\label{gainS}
\end{eqnarray}
where $\widetilde{p}_{e,v}(\theta_v,\sigma_v)=1-{p}_{e,v}(\theta_s,\sigma_s)$, $v\in\{b,\;u\}$ is the probability of perfect beam alignment at the BS and user. Since the spatial distribution of BSs follows a homogeneous PPP, the gain of the interfering links is also random. In particular, for a given interfering BS, the typical user is in the direction of the main-lobe gain with probability $\Hat{p}_b(\theta_b)=\frac{\theta_b}{2\pi}$. In the same manner, for the typical user, an interferer is in the direction of  the main-lobe gain with probability $\Hat{p}_u(\theta_u)=\frac{\theta_u}{2\pi}$. This implies that the gain between the typical user and any interfering BS is also a discrete r.v.  defined as\vspace{-0.2cm}
\begin{eqnarray}
     \Hat{G}(\theta_b,\theta_u)= \nonumber\\
&&\!\!\!\!\!\!\!\!\!\!\!\!\!\!\!\!\!\!\!\!\!\!\!\!\!\!\!\!\!\!\!\!\!\!\!\!\!\!\!\!\!\!\left\{
     \begin{array}{ll}
     \Hat{G}_{1}=G_{b}^MG_{u}^M, \;\;\textrm{w.p}\;\;\;  \Hat{p}_{1}= \Hat{p}_{b}(\theta_b)\Hat{p}_{u}(\theta_u)  \\
     \Hat{G}_{2}=G_{b}^SG_{u}^M, \;\;\;\textrm{w.p}\;\;\;  \Hat{p}_{2}= (1-\Hat{p}_{b}(\theta_b))\Hat{p}_{u}(\theta_u) \\
     \Hat{G}_{3}=G_{b}^MG_{u}^S, \;\;\;\textrm{w.p}\;\;\;  \Hat{p}_{3}= \Hat{p}_{b}(\theta_b)(1-\Hat{p}_{u}(\theta_u))\\
     \Hat{G}_{4}=G_{b}^SG_{u}^S, \;\;\;\;\textrm{w.p}\;\;\;  \Hat{p}_{4}= (1-\Hat{p}_{b}(\theta_b))(1-\Hat{p}_{u}(\theta_u)),
\end{array}
\right.
\label{GainI}
\end{eqnarray}\vspace{-0.3cm}
\subsection{Propagation Model}

Since the path loss and  molecular absorption in THz bands are extremely high, the connections rely mainly on the line-of-sight (LOS) links. Therefore, extremely high BS densities are anticipated for the deployment of THz wireless networks. This allows each user to be surrounded by multiple LOS BSs, making LOS connections dominant.
In this paper, we analyze the performance of a typical user located at the origin $(0,0)\in{\mathbb R}^{2}$  \cite{Baccelli2009stochastic}. The large-scale fading of a LOS link between the  user and a BS located at a distance $x$ is given by \cite{hossain2019stochastic} \vspace{-0.2cm}
\begin{eqnarray}
\Psi(x)=\frac{c^2}{(4\pi f)^2} x^{-\alpha}\exp{\big(-k_ax\big)},
\label{propmodel}
\end{eqnarray}
where $f$, $c$, and $\alpha$   denote the operating THz frequency, the speed of light, and the path loss exponent. In (\ref{propmodel}), $k_a$ represents the molecular absorption coefficient which depends on the operating frequency. Assuming the small-scale channel fading, denoted by $h$, follows a Nakagami distribution, then the channel gain  $\xi=|h|^2$ is a gamma r.v. with a PDF given as\vspace{-0.2cm}
\begin{eqnarray}
f_{\xi}(x)=\frac{\Omega^m}{\Gamma(m)}x^{m-1}e^{-\frac{1}{\Omega}x},
\label{chgain}
\end{eqnarray}
where $m$ and $\Omega$ are the shape and scale parameters, respectively, and $\Gamma(m)=\int_0^\infty t^{m-1} e^{-t}dt$ is the standard gamma function. Therefore,
the received signal power  from a serving BS at a distance y from the typical user is expressed as  $P_R(y)=P_TG(\theta_b,\theta_v,\sigma_b,\sigma_u)\Psi(y)$, where $P_T$ is the BS transmit power.  \vspace{-0.5cm} 
\subsection{BS Clustering Model}
In user-centric cooperative networks,  each user is served by a carefully selected group of BSs called the serving cluster and denoted here by ${\Phi}^{S}_{b}$. To select the user serving cluster, we used an adaptive clustering method as follows. 
Let $\tau=\underset{k\in \Phi_b}{\max}\; { \bar{P}}_R(r_k)$  be the maximum average power received by the typical user, where ${ \bar{P}}_R(r_k)$ is the average signal power received from the $k$-th BS located at a distance $r_k$. Here,  the BS with the maximum received power is referred to as the reference BS. Then, the BSs in the user's serving cluster are selected according to the following clustering model
\begin{equation}
{\Phi}^{S}_{b}=\{y_k\in \Phi_b,\;0 \leq\mid\mid y_k\mid\mid\leq R\},
\label{eqw}
\end{equation}
where $R=\frac{r}{\delta}$ is the radius of the serving cluster with  $r$ being the distance between the user and the BS providing the maximum  power and is obtained as 
$r=\frac{\alpha}{k_a} \textrm{W}_0\big(\alpha^{-1}k(f) \big(\frac{\tau}{CG^M_bG^M_u}\big)^{\frac{-1}{\alpha}}\big)$, 
where $\textrm{W}_0(.)$ is the Lambert W$_0$ function defined as the inverse of the function $f(w)=we^{w}$ and $C=\frac{c^2}{(4\pi f)^2}$. The parameter, $\delta \in [0, 1]$ is called the clustering parameter, used to control the cluster size. 
 The clustering model in (\ref{eqw}) implies that different users may have  different cluster sizes. Moreover, if any user updates its position, the  cluster size as well as serving BSs within the cluster may change. 


\section{User-Centric Coverage  Analysis}
In this section, we begin with developing  mathematical definitions for
 the signal-to-interference-plus-noise ratio (SINR)  as well as  the stochastic  distribution of the  interference at the typical user. Then, we derive a theoretical expression of the coverage probability under the proposed adaptive clustering model in the presence of beam alignment error. \vspace{-0.35cm}
\subsection{Received SINR} 
The
received SINR  at the
typical user, denoted as $\Upsilon$,  can be defined as\vspace{-0.3cm}
\begin{eqnarray}
\Upsilon=\frac{A+B}{I+\sigma^2_n}  
  \label{eq7}
\end{eqnarray}
where $\sigma^2$
is the thermal noise power, $A$ is  the  signal power  from the reference BS, $B$  is the aggregated signal  from the other serving BSs,  and $I$ is the aggregated interference.  The quantities $A$, $B$, and $I$  are given as 
\begin{eqnarray}
A=P_TG(\theta_b,\theta_u,\sigma_b,\sigma_u)\xi {\Psi}(R\delta),
\label{refBS}
\end{eqnarray} \vspace{-0.3cm}
\begin{eqnarray}
B=\sum_{i\in {\Phi}^{S}_b\setminus{\cal B}(0,R\delta)}P_TG_i(\theta_b,\theta_u,\sigma_b,\sigma_u)\xi_i {\Psi}(r_i)
\label{agrrSig},
\end{eqnarray}
and \vspace{-0.35cm}
\begin{eqnarray}
I= \sum_{j\in \Phi_{b}\setminus {\Phi}^{S}_b}P_T\Hat{G}_j(\theta_b,\theta_u)\xi_j {\Psi}(r_j),
\label{agrrInt}
\end{eqnarray}
where ${\cal B}(0,R\delta)$ denotes a ball of radius $R\delta$, $G_k(.)$ and $\Hat{G}_j(.)$ define the tatal gain of the links from the typical user to the $k$-th serving BS and to the $j$-th interfering BSs, respectively. 
 The  distance from the typical user to the $k$-th serving BS located at $y_k$ $\in$ $\Phi_b^s$ is denoted by $r_k$ while the distance from the $j$-th interfering BS located at $y_j$  $\in$  $\Phi_b\!\setminus\!{\Phi}^{S}_b$ is denoted by $r_j$.  Due to the extremely high sensitivity to blockages, THz BSs are expected to be spatially distributed with ultra-high density which makes the LOS links dominant. Therefore, in this work, we consider only the LOS links in the analysis. 
To add the effect of blockages on the received THz signals, we imply the Boolean blockage model \cite{bai2014coverage}. Using this model, the $k$-th serving BS located at a distance $r_k$ from the typical user is considered LOS with probability $p_L(r_k)=e^{-\beta r_k}$, where $\beta$ is a constant reflecting the blockage density and distribution.

\vspace{-0.3cm}
\subsection{Coverage Probability}
In this subsection, we provide mathematical expression for the SINR coverage probability of the THz user-centric network under beam alignment error. Before conducting the coverage probability derivations, we first compute the distribution of the aggregated interference. 
The following lemma calculates the  characteristic functions (CFs) of the aggregated signal and aggregated interference.

\textit{\bf Lemma 1}: For a given cluster radius, $R$, the CF of the aggregated signal $B$ denoted by $\Lambda_{B|R}(jw)$ and the CF of the aggregated interference $I$  denoted by  $\Lambda_{I|R}(jw)$ are, respectively, given in (\ref{SigT}) and (\ref{LTi}) at the top of the next page.
\begin{figure*}
\begin{eqnarray}
\Lambda_{B|R}(j\omega)=\mathbb{E}_\xi\Bigg[\prod_{n=1}^\infty \exp\Bigg(-2\pi\lambda_by\sum_{i=1}^4 \frac{p_i(-jwCG_i P_T\xi)^{n}\Big(\Gamma(2-\alpha n,\;n(k_a+\beta)R\delta)-\Gamma(2-\alpha n,\;n(k_a+\beta)R)\Big)}{n!(nk_a+n\beta)^{2-\alpha n}}\Bigg)\Bigg].
\label{SigT}
\end{eqnarray}
\begin{eqnarray}
\Lambda_{I|R}(j\omega)=\mathbb{E}_\xi\Bigg[\prod_{n=1}^\infty\exp\Bigg(-2\pi\lambda_b\sum_{j=1}^4 \frac{\Hat{p}_j(-sC\Hat{G}_j P_T\xi)^{n}\Gamma(2-\alpha n,\;n(k_a+\beta)R)}{n!(nk_a+n\beta)^{2-\alpha n}}\Bigg)\Bigg].
\label{LTi}
\end{eqnarray}
\hrulefill
\end{figure*}
\textit{Proof:} 
From (\ref{agrrSig}), the CF of the aggregated signal $B$ for a given $R$, is expressed as
\begin{eqnarray}
\Lambda_{B|R}(jw)&=&\mathbb{E}_{B}\big[e^{-jw B}\big]\nonumber\\
&&\!\!\!\!\!\!\!\!\!\!\!\!\!\!\!\!\!\!\!\!\!\!\!\!\!\!\!\!\!\!\!\!\!\!\!\!\!\!\!\!\!\!=\mathbb{E}_{G,\Phi_b^S,\xi} \Bigg[\exp{\Bigg(\!\!\!\!-jw\!\!\!\!\!\!\!\! \sum_{i\in {\Phi}^{S}_b\setminus{\cal B}(0,R\delta)}\!\!\!\!\!\!\!\!P_TG_i(\theta_b,\theta_u,\sigma_b,\sigma_u)\xi_i \Psi (r_i)\Bigg)}\Bigg]\nonumber\\
&&\!\!\!\!\!\!\!\!\!\!\!\!\!\!\!\!\!\!\!\!\!\!\!\!\!\!\!\!\!\!\!\!\!\!\!\!\!\!\!\!\!\!\overset{(a)}{=}\mathbb{E}_{\xi} \Bigg[\exp\Bigg(\!\!\!-2\pi\lambda_b\sum_{i=1}^4 p_i\int_{\delta R}^R \!\!\!\big(1\!\!-\!e^{-jwG_i  P_T C\xi t^{-\alpha} e^{-k_a t}}\nonumber\\
&&\times te^{-\beta t}dt\big)\Bigg)\Bigg]\nonumber\\
&&\!\!\!\!\!\!\!\!\!\!\!\!\!\!\!\!\!\!\!\!\!\!\!\!\!\!\!\!\!\!\!\!\!\!\!\!\!\!\!\!\!\!\overset{(b)}{=}\mathbb{E}_{\xi} \Bigg[\exp-2\pi\lambda_b\sum_{i=1}^4 p_i\int_{\delta R}^R \sum_{n=1}^\infty \frac{(-jwG_iP_T\xi C)^n}{n!}\nonumber\\
&&\times \frac{t^{-n\alpha+1} e^{-n (k_a+\beta)t}}{n!} dt\Bigg]
\label{app1}
\end{eqnarray}
where (a) results from using the probability generating
functional of a PPP \cite{Baccelli2009stochastic} and from taking the average over the gain in (\ref{gainS}), (b) follows from expanding inner exponential function. Then, applying $\int_{z_1}^{z_2}  t^{a-1}e^{-t}dt=\Gamma(a,z_2)-\Gamma(a,z_1)$, $z_2>z_1$, yields the expression in (\ref{SigT}). Similarly, the results in (\ref{LTi}) can be derived by following the same steps.

From (\ref{SigT}) and (\ref{LTi}), one can observe that it is infeasible to calculate the SINR coverage using characteristic functions-based methods, e.g. Gil-Pelaez inversion theorem \cite{di2014stochastic}, because of the infinite multiplication in (\ref{SigT}) and (\ref{LTi}).
As such, we imply the central limit theorem (CLT) to approximate the distribution of the aggregated interference with a
normal distribution. 
Therefore, the following proposition computes the conditional  distribution of the aggregated interference.

\textbf{Proposition 1:} For a give cluster radius $R$, the CDF of the aggregated interference is approximated as 
\begin{eqnarray}
F_{I|R}(T)\!\!\!\!\!\!&=\!\!\!\!\!\!&\textrm{Pr}\Big(\sum_{j\in \Phi_{b}\setminus {\Phi}^{S}_b}P_T\Hat{G}_j(\theta_b,\theta_u)\xi_j {\Psi}(r_j)\leq T\Big)\nonumber\\
&&\!\!\!\!\!\!\approx \frac{1}{2}\bigg(1+\textrm{erf}\bigg(\frac{T-\mu_{I}(R)}{\sqrt{2 \textrm{var}_{I}(R)}}\bigg)\bigg),
\label{cdfi}
\end{eqnarray}
where $\textrm{erf}(z)=\frac{2}{\sqrt{\pi}}\int_0^ze^{-t^2}dt$ is the error function \cite{gradshteyn2014table}, and  $\mu_{I}(R)$ and $\textrm{var}_{I}(R)$ are, respectively, the mean and variance of the aggregated interference, $I$, for a given value of $R$ which are obtained as 
$\mu_{I}(R)={\cal M}^{(1)}(R)$
$\textrm{var}_{I}(R)={\cal M}^{(2)}(R)-\big({\cal M}^{(1)}(R)\big)^2$, 
where ${\cal M}^{(1)}(R)$ and ${\cal M}^{(2)}(R)$ are the mean and the mean-squared values of  $I$ for a given $R$ which can be obtained  by taking the first and second derivatives of the CF in (\ref{LTi}). Both ${\cal M}^{(1)}(R)$ and ${\cal M}^{(2)}(R)$ are expressed in (\ref{mom1}) and (\ref{mom2}) at the top of the next page,  
\begin{figure*} \vspace{-0.35cm}
\begin{eqnarray}
{\cal M}^{(1)}(R)&=&-\frac{\partial \Lambda_{I|R}(jw)}{j\partial w}\bigg|_{w=0}=\frac{2\pi \lambda_b C \Hat{G}^{(1)}m\Gamma\big(2-\alpha,\;(k_a+\beta)R\big)}{\Omega(k_a+\beta)^{2-\alpha}},
\label{mom1}
\end{eqnarray}
\hrulefill
\end{figure*}
\begin{figure*}\vspace{-0.4cm}
\begin{eqnarray}
\!\!\!\!{\cal M}^{(2)}(R)&\!\!\!\!\!\!=&\!\!\!\!\!\!\frac{\partial^2 \Lambda_{I|R}(jw)}{j^2\partial w^2}\bigg|_{w=0}\!\!\!\!\!\!\!\!=\!\!\frac{\big(2\pi \lambda_b C \Hat{G}^{(2)} m (1 + m)\Gamma\big(2-\alpha,\;(k_a+\beta)R\big)\big)^2}{\Omega^2(k_a+\beta)^{4-2\alpha}}\!+\!\frac{4\pi \lambda_b C^2 \Hat{G}^2m^2\Gamma\big(2-2\alpha,\;2(k_a+\beta)R\big)}{2\Omega^2\big(2(k_a+\beta)\big)^{2-2\alpha}}.
\label{mom2}
\end{eqnarray}
\hrulefill
\end{figure*}
where $\Hat{G}^{(1)}$ and $\Hat{G}^{(2)}$ are the first and second moments of the gain in (\ref{GainI}), respectively.

It is worth mentioning that according to \cite{elsawy2016modeling,aljuaid2010investigating}, using the CLT for normal approximation of the interference distribution becomes more efficient with increasing the density and when the interference components with dominant power are excluded. In other words, the accuracy of using the CLT diminishes with low-dense networks.  
In the proposed BS clustering scheme, the user's serving cluster can be considered an interference exclusion region. For this reason and since THz BSs are expected to be spatially distributed with ultra-high density, the aggregate interference is a sum of a massive number of independent r.vs. with comparable variances, which demonstrates the validity of using  the CLT.

To proceed, we need to compute the distribution of the cluster radius  $R$ as given in the following lemma

\textit{\bf Lemma 2:} Given the  clustering parameter $\delta$, the PDF of the serving cluster radius, $R$,  is expressed as
\begin{eqnarray} 
  f_{R}(R)&\!\!\!\!\!=&\!\!\!\!\frac{2\pi\lambda_b \delta R }{e^{\beta \delta R +2\pi\lambda_b \Big(\beta^{-2}\!-\!\beta^{-1}\delta R e^{-\beta \delta R}\!-\!\beta^{-2}e^{-\beta \delta R}\Big)}},
  \label{eqr}
\end{eqnarray}
\textit{Proof:} Since the user's serving cluster is assumed to contain only LOS BSs, the distance to the reference BS, $r$, is equivalent to the distance to the closest LOS BS which follows a Rayleigh distribution in the case of PPP BS distribution with a density $\lambda$ scaled by the LOS probability $p_L(r)=e^{-\beta r}$. Then, using  the cluster radius $R=\delta^{-1} r$, the PDF of $R$ can be written as in  (\ref{eqr}). 

 \textbf{Proposition 2:}
 Given  the SINR threshold, $\gamma$, the coverage probability of the proposed user-centric THz wireless network under beam alignment error denoted as ${\cal C}(\gamma)$   is given  in (\ref{covProb}) at the top of the  page,
\begin{figure*}\vspace{-0.3cm}
\begin{eqnarray}
{\cal C}(\gamma)= \int_0^\infty\frac{1}{2}\Bigg(1+\sum_{q=0}^\infty \sum_{i=1}^4p_i\textrm{erf}\Bigg(\frac{P_TG_i C_T (R\delta)^{-\alpha} e^{-k_a(f)R\delta}
+q D(R)- \big(\sigma^2+\mu_{I_{agg}}(R)\big)}{\gamma \sqrt{2 \textrm{var}_{I_{agg}}(R)}}\Bigg)\Bigg)f_Q(q|R)f_R(R)dR,
\label{covProb}
\end{eqnarray}
\hrulefill
\end{figure*}
where $f_Q(q|R)$ is the pmf of the random variable Q given $R$ which is expressed as
\begin{eqnarray}
f_Q(q|R)=\frac{(\pi \lambda R^2(1-\delta^{2}))^q}{q!} e^{-\pi \lambda R^2(1-\delta^{2})},
\end{eqnarray}
and $D(R)$ is given as
\begin{eqnarray}
D(R)&=&2P_TC_T {G}^{(1)}m\Omega^{-1} k_a^{-\alpha+2}(\delta R)^{-2}(\delta^{-2}-1)^{-1}\nonumber
\\
&&\times\Big(\Gamma\big(2-\alpha,k_a \delta R\big)-\Gamma\big(2-\alpha,k_a R\big)\Big),
\end{eqnarray}
where $G^{(1)}$ is the first moment of the gain in (\ref{gainS}).
\\
  \textit{Proof:} See the Appendix.

  From (20), we can observe that for large cluster sizes, i.e, $\delta\rightarrow0$, the argument of $\textrm{erf}(.)$ goes to infinity and hence ${\cal C}(\gamma)\rightarrow 1$. This corresponds to a full-cooperation case where all BSs contribute to serving the user, i.e., no interference. Although this may be feasible in small-scale networks, it becomes impractical as the size of the network increases. Therefore,  a user-centric BS cooperation is introduced, in which each user is connected to a subset of BSs instead of all BSs in the large-scale networks.
\vspace{-0.3cm}
\section{Numerical Results}
In this section, we present a numerical study of the introduced user-centric THz networks with beam alignment error  using the analytical expressions from Section III and Monte Carlo (MC) simulations based on the system model developed in Section II.
Unless otherwise specified, the simulation parameters are considered as follows. The BS transmit power is set to $P_T=30$ dBm, the path-loss exponent is $\alpha=2.5$, and the BS density is $\lambda_b=0.005 \;\textrm{BSs/m}^2$.
 The absorption coefficient is $0.06$ which corresponds to an operating frequency of  $1$ THz, based on the practical data collected for water vapor molecules absorption with humidity $36.78\%$  \cite{slocum2013atmospheric}. The number of antenna elements in BSs and users is $N_b=N_u=8$  and the blockage parameters is $\beta=1/141.4$.
 
\begin{figure}[t]
    \centering
    \includegraphics[scale=0.65]{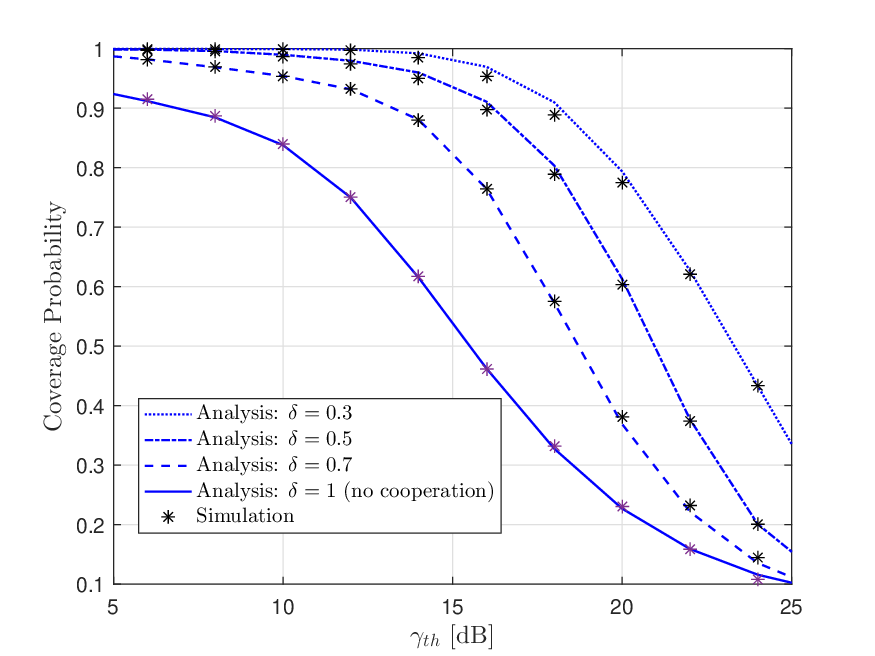}
    \caption{ Coverage probability versus the SINR threshold $\gamma$ for   different values of the clustering parameter with $\sigma_b=\sigma_u=10^\circ$}
    \label{fig1}
    \vspace{-0.5cm}
    \end{figure}
    
    \begin{figure}[t]
    \centering
    \includegraphics[scale=0.65]{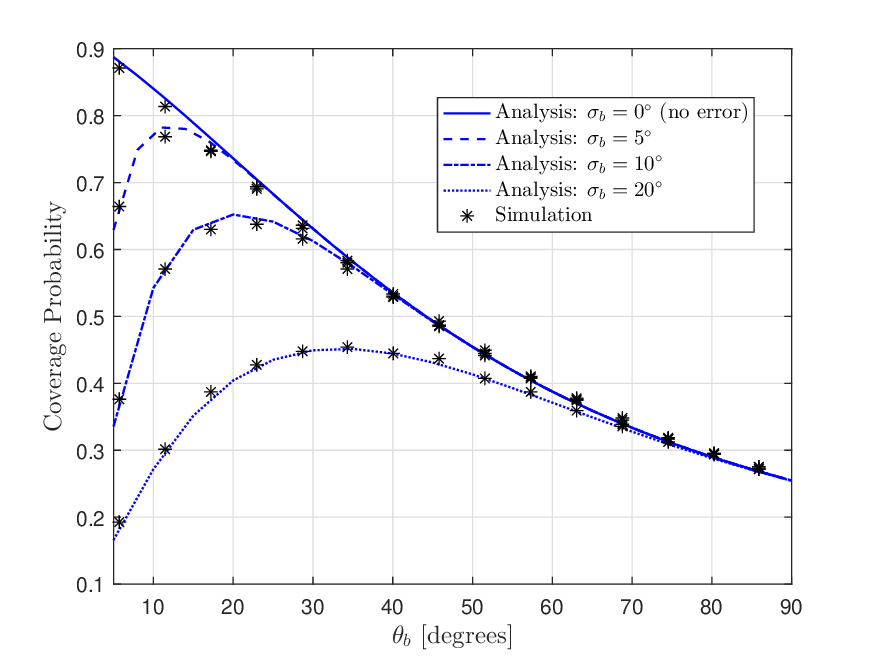}
    \caption{ Coverage probability versus BS antenna beamwidth $\theta_b$ for different values of the standard deviation of the alignment error in BSs, $\sigma_b$,  with $\sigma_u=10^\circ$.}
    \label{fig2}
    \vspace{-0.5cm}
    \end{figure}
    
    \begin{figure}[t]
    \centering
    \includegraphics[scale=0.65]{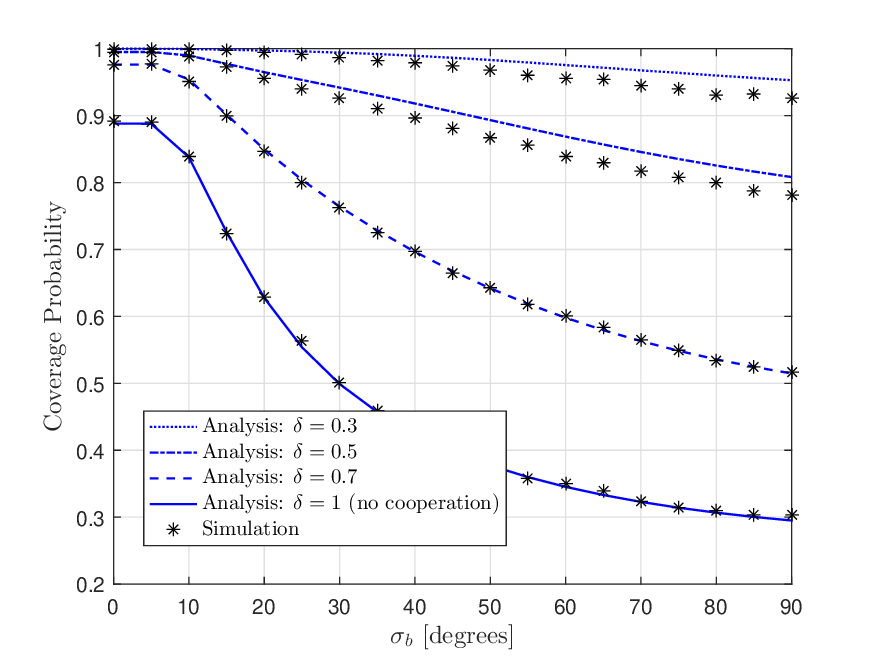}
    \caption{ Coverage probability versus standard deviation of the alignment error in BSs, $\sigma_b$, for   different values of the clustering parameter with $\sigma_u=10^\circ$.}
    \label{fig3}
    \vspace{-0.5cm}
    \end{figure}

Fig. \ref{fig1} reveals the impact of increasing the cooperative cluster size (decreasing $\delta$), in the presence of beam misalignment with $\sigma_b=\sigma_u=10^{o}$. We  observe that decreasing $\delta$ results in a higher coverage probability 
due to a larger macro-diversity gain via the cooperation of larger number of BSs. The coverage improvement due to user-centric  BS cooperation  is significant compared to  that
of  single-BS connection, i.e. non-cooperation when $\delta=1$. 
We observe that the analytical expression
of the coverage probability leveraging the CLT-based approximation of the interference distribution in (17) is in close agreement
with MC simulations. 


Fig. \ref{fig2} demonstrates  the impact of varying the BS antenna beamwidth $\theta_b$ for different values of the beam misalignment severity. It can be observed that in the absence of beam misalignment,  decreasing $\theta_b$ improves the coverage probability performance since using antenna arrays with narrow beams significantly decrease the interference at  the user. In the presence of beam misalignment, it can be seen that  the coverage probability  improves initially with decreasing $\theta_b$ until it reaches a maximum value and then starts decaying, giving an optimum beamwidth.    
 This results from the fact that narrow beams are more sensitive to the beam misalignment error which degrades the received signal power and increases the  interference. It can also be noted that the optimal value of $\theta_b$ increases with $\sigma_b$.

 Fig. \ref{fig3} plots the coverage probability as a function of the beam misalignment severity $\sigma_b$.
 This figure shows that for non-cooperative transmission (i.e $\delta=1$),  the coverage probability  drastically decreases when  $\sigma_b$ increases. On the other hand, applying user-centric BS cooperation mitigates the effect of the beam misalignment, which significantly decreases with increasing the cooperation cluster size, i.e., decreasing $\delta$.  As such, Fig. \ref{fig3} clearly shows that user-centric clustering using joint transmission from a large cluster of BSs is a useful add-on to THz networks in the presence of beam misalignment.

\begin{figure}[t]
    \centering
    \includegraphics[scale=0.65]{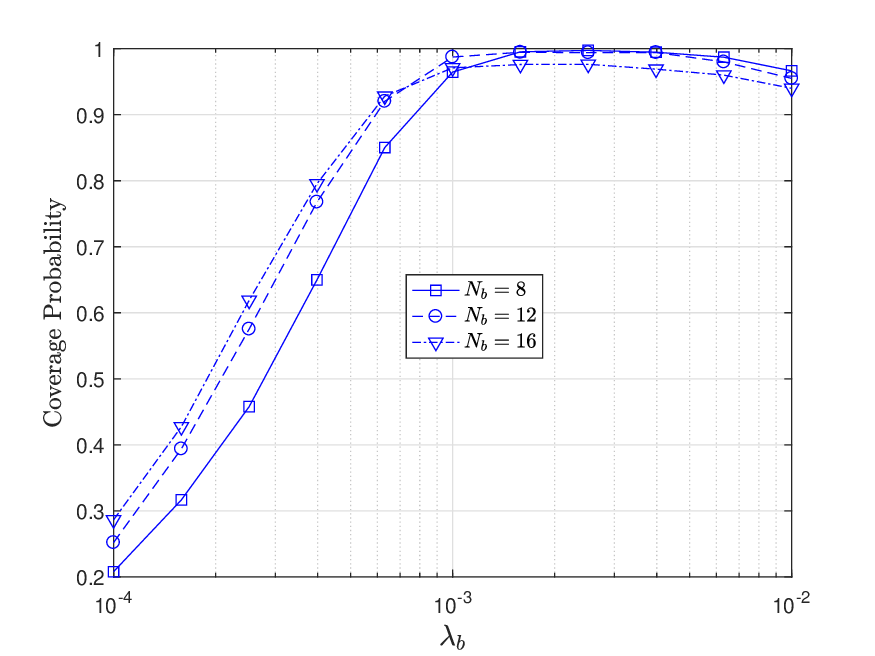}
    \caption{Coverage probability versus BS density $\lambda_b$ for different BS antenna array configurations with  $\delta=0.6$ $\sigma_b=\sigma_u=5^\circ$, $N_u=8$.
        }
    \label{fig4} 
    \vspace{-0.5cm}
\end{figure}

In Fig. \ref{fig4}, we plot the coverage probability as a function of BS density, $\lambda_b$ for three BS antenna array configurations, i.e., $N_b=8, 12$ and $16$. As shown in this figure, increasing the antenna array size can enhance the system performance at low densities, but it slightly degrades the performance at high BS density deployments. 
This occurs because the increase in the number of antenna elements not only decreases the antenna beamwidth but also heightens sensitivity to beam misalignment, resulting in increased interference.
This interference becomes more harmful at ultra-dense deployments.  Furthermore, Fig. \ref{fig4} demonstrates that 
 coverage probability increases with the BS density to a specific
point and starts degrading afterward. This implies that there is always 
a critical BS density which provides maximum performance. 

Based on the provided numerical results, several system-level insights and design guidelines can be derived for user-centric THz networks with beam alignment errors. First, the coverage improvement due to user-centric BS cooperation is substantial, particularly when compared to single-BS connections. However, this increases the fronthauling overhead and complexity. In addition, due to beam misalignment errors, there is always an optimum value of the antenna beamwidth, $\theta_b$. Beyond this value, further decreasing $\theta_b$ degrades the system performance, where narrow beams are more sensitive to misalignment errors, which degrade received signal power and increase interference. Furthermore, increasing the antenna array size can enhance system performance at low BS densities but may slightly degrade the performance at high BS density deployments. This is because increasing the antenna arrays size decreases the antenna beamwidth, which can heighten sensitivity to beam misalignment and increase interference, particularly in ultra-dense deployments.
\section{Conclusion}

In this letter, we  introduced a stochastic geometry framework to investigate the coverage performance of
user-centric THz networks by considering inherent beam misalignment. The system performance
in terms of coverage probability is analysed by leveraging a unified CLT-based   characterization of the aggregate interference statistics.  Besides validating our theoretical
finding, the numerical results assess the impact of beam 
misalignment with respect
to crucial link parameters, such as
the transmitter’s beam width and  the serving  cluster size, and also demonstrate that user-centric BS clustering can mitigate the effect of beam misalignment in THz networks.\vspace{-0.3cm}
\section{Appendix}
\subsection*{Proof of Proposition 2}
Given the cluster radius is $R$ and the SINR threshold is $\gamma$, the coverage probability is defined as\vspace{-0.3cm}
\begin{eqnarray}
{\cal C}(\gamma)&=&\textrm{Pr}(\Upsilon>\gamma)=\textrm{Pr}\bigg(I<\frac{A+B-\gamma \sigma^2}{\gamma}\bigg)\nonumber\\
&&\!\!\!\!\!\!\!\!\!\!\!\!\!\!\!\!\!\!\!\!\!\!\!\!\!\!\!\!\!\!\!\!\!=\mathbb{E}_{A,B,R}\Bigg[ F_{I|r}\bigg(\frac{A+B-\gamma \sigma^2}{\gamma}\bigg)\Bigg]\nonumber\\
&&\!\!\!\!\!\!\!\!\!\!\!\!\!\!\!\!\!\!\!\!\!\!\!\!\!\!\!\!\!\!\!\!\!\overset{(a)}{=}\mathbb{E}_{A,B,R}\Bigg[\frac{1}{2}\Bigg(1+\textrm{erf}\bigg(\frac{A+B-\gamma \sigma^2-\gamma\mu_{I}(R)}{\gamma\sqrt{2 \textrm{var}_{I}(R)}}\bigg)\Bigg)\!\!\Bigg],
\label{sinrp}
\end{eqnarray}
where (a) follows from recalling (\ref{cdfi}) in Proposition 1. 
For the aggregated signal $B$ in (\ref{agrrSig}), the $k$-th distance $r_k$, is a r.v. uniformly distributed in $[\delta R,\; R]$ with a PDF
$f_{r_k}(t)=\frac{2t}{(\delta R)^2(\delta^{-2}-1)}$.
Since BSs are a PPP distributed with density $\lambda$, the number of $Q$ BSs in ${\Phi}^{S}_b\setminus {\cal B}(0,\delta R)$ represents a Poisson distribution with a mean $\pi\lambda_b (\delta R)^2(\delta^{-2}-1)$ and a PMF given  as 
$\textrm{Pr}[Q=q]=\frac{\big(\pi\lambda_b (\delta R)^2(\delta^{-2}-1)\big)^q}{q!}e^{-\pi\lambda_b (\delta R)^2(\delta^{-2}-1)}$.
As such, the coverage probability in (\ref{sinrp}) can be expressed as
\begin{eqnarray}
\textrm{P}_{cov}^d(\gamma)&=&\nonumber\\
&&\!\!\!\!\!\!\!\!\!\!\!\!\!\!\!\!\!\!\!\!\!\!\!\!\!\!\!\!\!\!\!\!\!\!\!\!\!\!\!\!\!\mathbb{E}_{A,Q,R}\Bigg[\frac{1}{2}\Bigg(1+\textrm{erf}\bigg(\frac{A+QD(R)-\gamma \sigma^2-\gamma\mu_{I}(R)}{\gamma\sqrt{2 \textrm{var}_{I}(R)}}\bigg)\Bigg)\!\!\Bigg],
\label{App20}
\end{eqnarray}
where $F(R)$ is given by\vspace{-0.3cm}
\begin{eqnarray}
D(R)&=& P_TC_T {G}^{(1)}m\Omega^{-1}\int_{\delta R}^{R}t^{-\alpha}e^{-k_at} f_{r_k}(t)dt\nonumber\\
&&\!\!\!\!\!\!\!\!\!\!\!=2P_TC_T {G}^{(1)}m\Omega^{-1} k_a^{-\alpha+2}(\delta R)^{-2}(\delta^{-2}-1)^{-1}\nonumber
\\
&&\times\Big(\Gamma\big(2-\alpha,k_a \delta R\big)-\Gamma\big(2-\alpha,k_a R\big)\Big).
\label{App21}
\end{eqnarray}
Then, the expression in (\ref{covProb}) can be obtained by plugging (\ref{refBS}) and (\ref{App21}) in (\ref{App20}) and averaging over $Q$, the gain in (\ref{gainS}), and the cluster radius $R$ in (\ref{eqr}).


\vspace{-0.3cm}
\bibliographystyle{IEEEtran}
\def\bibfont{\footnotesize}
\bibliography{References2.bib}






\end{document}